\documentclass[pre,longbibliography,superscriptaddress]{revtex4-2}

\usepackage{amsmath,amssymb,latexsym,epsfig,graphics,epsf,bm}
\usepackage{xcolor}
\usepackage{graphics,xcolor}
\usepackage{float}
\usepackage[a4paper, textwidth=180mm,textheight=240mm]{geometry}

\newcommand{\eq}[1]{Eq.~(\ref{#1})}

\newcommand{\be}{\begin{equation}}
\newcommand{\ee}{\end{equation}}

\newcommand{\fig}[1]{Fig.~\ref{#1}}
\newcommand{\Fig}[1]{Figure~\ref{#1}}

\begin{document} 

\title{Time-reversibility during the ageing of materials}

\author{Till B{\"o}hmer}\affiliation{Institute for Condensed Matter Physics, Technical University of Darmstadt, D-64289 Darmstadt, Germany}
\author{Jan P. Gabriel}\affiliation{{\it Glass and Time}, IMFUFA, Department of Science and Environment, Roskilde University, P. O. Box 260, DK-4000 Roskilde, Denmark} 
\author{Lorenzo Costigliola}\affiliation{{\it Glass and Time}, IMFUFA, Department of Science and Environment, Roskilde University, P. O. Box 260, DK-4000 Roskilde, Denmark}
\author{Jan-Niklas Kociok}\affiliation{Institute for Condensed Matter Physics, Technical University of Darmstadt, D-64289 Darmstadt, Germany} 
\author{Tina Hecksher}\affiliation{{\it Glass and Time}, IMFUFA, Department of Science and Environment, Roskilde University, P. O. Box 260, DK-4000 Roskilde, Denmark} 
\author{Jeppe C. Dyre}\affiliation{{\it Glass and Time}, IMFUFA, Department of Science and Environment, Roskilde University, P. O. Box 260, DK-4000 Roskilde, Denmark}
\author{Thomas Blochowicz}\affiliation{Institute for Condensed Matter Physics, Technical University of Darmstadt, D-64289 Darmstadt, Germany}

\date{\today}

\begin{abstract}
Physical ageing is the generic term for irreversible processes in glassy materials resulting from molecular rearrangements. One formalism for describing such ageing processes involves the concept of material time, which may be thought of as time measured on a clock whose rate changes as the glass ages. Experimental determination of material time has so far not been realized, however. Here, we show how dynamic light-scattering measurements provide a way forward. We determined the material time for an ageing sample of the glass-former 1-phenyl-1-propanol after temperature jumps close to the glass transition from the time-autocorrelation function of the intensity fluctuations probed by multispeckle dynamic light scattering. These fluctuations are shown to be stationary and reversible when regarded as a function of the material time. The glass-forming colloidal synthetic clay Laponite and a chemically ageing curing epoxy are also shown to display material-time-reversible scattered-light intensity fluctuations, and simulations of an ageing binary system monitoring the potential energy confirm material-time reversibility. In addition to demonstrating direct measurements of the material time, our findings identify a fundamental property of ageing in quite different contexts that presents a challenge to the current theories of ageing.
\end{abstract}

\maketitle

\begin{figure}
	\centering
	\includegraphics[width=60mm]{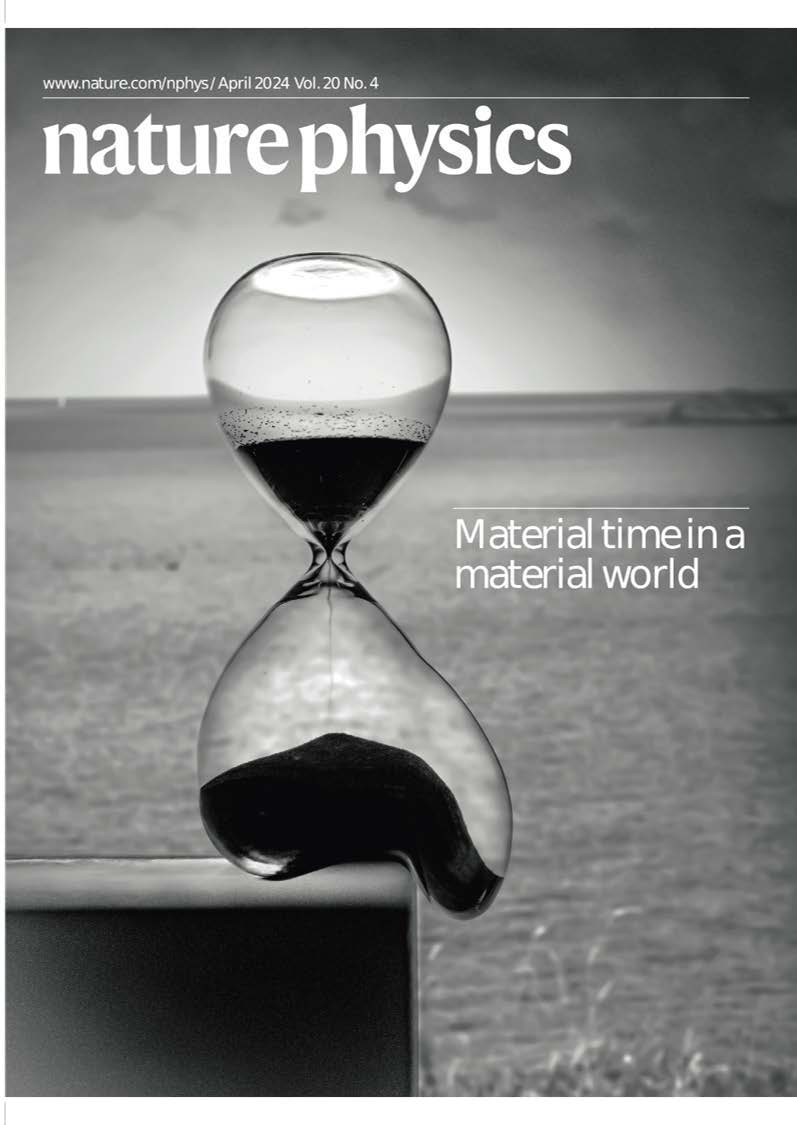}
\end{figure}

\maketitle

\section{Introduction}\label{intro}

Few persons would claim that time can be reversed -- all living creatures age and eventually die, a dropped glass breaks while the reverse never happens, mixing cold and warm water leads irreversibly to an in-between temperature and so on \cite{rovelli}. On the other hand, the fundamental equation-of-motion laws of nature are all time reversible: for example, Newton’s laws of classical mechanics, the Schr\"odinger equation of quantum mechanics, Maxwell’s equations for electromagnetism and the Einstein equation of gravity and space-time. The irreversibility of everyday life is accounted for by the second law of thermodynamics, which states that the entropy of an isolated system can increase or stay constant but never decrease \cite{lan58,reichl}. 

When entropy is constant in time, the system in question is in thermal equilibrium. Reflecting the time reversibility of the fundamental laws, thermal-equilibrium fluctuations are statistically time reversible. As shown by Onsager in 1931, this reversibility leads to quantitative relations between coefficients describing small, linear deviations from equilibrium \cite{ons31}. This paper presents ageing data that go much beyond the linear-response regime, but are nevertheless shown to be statistically reversible when regarded as functions of the so-called material time that controls ageing according to the classical Tool-Narayanaswamy (TN) formalism \cite{too46,nar71}. This goes against the conventional wisdom according to which ageing is fundamentally irreversible, involving plasticity, dissipation, entropy production, and dynamical heterogeneities with a correlation length that changes during ageing. Our findings from light-scattering experiments on three quite different samples, confirmed by computer simulations of a simple model system, bring into question the status of ageing as a prototypical case of non-equilibrium thermodynamics \cite{man21}.

\section{Physical ageing}\label{pa}
In contrast to degradation involving chemical reactions as in corrosion, physical ageing involves changes of material properties that are caused exclusively by molecular rearrangements \cite{str78,scherer,hod95,che07,mic16,mck17,rut17,can18,arc20}. Non-crystalline materials like ordinary glass \cite{nar71,scherer}, polymers \cite{mck94,hod95,mck17}, and metallic glasses \cite{rut17,mon20,zha22}, are all subject to physical ageing because the glassy state relaxes continuously toward a state of metastable equilibrium \cite{sim31}. In the vast majority of cases this is too slow to be observed, but in certain cases physical ageing results in undesirable property changes. The study of physical ageing is important for applications of glassy materials, as well as for optimizing their production. For this reason -- and because what controls the rate of physical ageing remains disputed -- this old research field continues to attract attention \cite{roth,lul20,man20,pas21,jan22,sch21,Elizondo2022}.

The best controlled physical ageing experiments involve a rapid change of temperature starting from a state of (metastable) thermal equilibrium, after which the system’s path toward equilibrium at a fixed ``annealing’’ temperature is monitored by continuously measuring some physical property. The outcome of such a temperature-jump experiment depends critically on whether a jump up or down is considered. Comparing up and down jumps to the same temperature, the latter are significantly faster and more stretched in time than up jumps \cite{scherer,mck17}. Even jumps involving just a one percent temperature change result in quite different relaxations toward equilibrium, in some cases with more than a decade difference in average relaxation time. This ``asymmetry of approach'' implies that physical ageing is highly nonlinear \cite{scherer,kob00,mck17}.

\section{Experimental methods in brief}
\begin{figure}
	\centering
	\includegraphics[width=180mm]{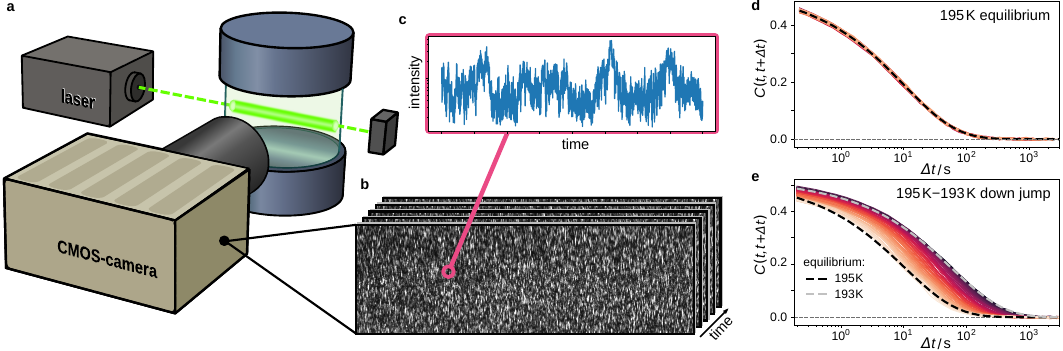}
	\caption{\textbf{Multispeckle dynamic light scattering.} 
		(a) The sample is illuminated by a laser beam, and an sCMOS-camera is used to detect the speckle pattern (b) of the 90$^\circ$ scattered-light intensity. 
		(c) In each speckle statistically independent temporal intensity fluctuations are observed. The time-resolved normalized intensity time-autocorrelation function, $C(t,t+\Delta t)$, is obtained as the multi-speckle average. 
		(d) Data for 1-phenyl-1-propanol in thermal equilibrium at 195\,K where $C(t,t+\Delta t)$ is stationary, i.e., depends only on $\Delta t$. 
		(e) During physical ageing following a temperature down jump, $C(t,t+\Delta t)$ changes with $t$ for fixed $\Delta t$. The evolution starts from the equilibrium time-autocorrelation function at the initial temperature $195\,$K (left dashed curve) and ends in equilibrium at the annealing temperature $193\,$K (right dashed curve). The different colors correspond to different times $t$ evenly spaced on a logarithmic axis.}
	\label{fig1}
\end{figure}

In many experiments physical ageing is well described in terms of the so-called material time \cite{scherer}. This concept was introduced in 1971 by Narayanaswamy, who, as an engineer at Ford Motor Co., sought the cooling protocol for generating optimal frozen-in stresses in the production of windshields \cite{nar71}. The material time may be thought of as time measured on a clock the rate of which changes as the glass ages. Conceptually, this is analogous to the proper-time concept of the theory of relativity, the time measured on a clock following a moving observer. The material-time approach to physical ageing is termed the TN formalism \cite{scherer,mck17}, an equivalent of which was developed a few years later for ageing polymers \cite{kov79}. The striking mathematical simplification of TN is that physical ageing becomes \textit{linear} when described in terms of the material time. Thus for all temperature jumps, a given quantity relaxes toward equilibrium following the \textit{same} function of the material time, except for an amplitude scaling in proportion to the jump magnitude. 

Specifically, when the temperature history is parameterized in terms of the material time $\xi$, the TN formalism describes ageing by a linear convolution integral over the temperature-variation history \cite{nar71,scherer,mck94}. Using instead the laboratory time as parameter, the linear limit is only approached for temperature variations in the mK range. The fundamental prediction of TN is that knowledge of this linear limit is enough to determine the highly nonlinear ageing phenomenon. This was recently verified in a paper reporting temperature jumps of amplitude down to 10\,mK \cite{rie22}.

According to the TN formalism, linear-response theory applies for the material-time development of average quantities. By the fluctuation-dissipation theorem \cite{reichl}, a linear response reflects equilibrium thermal fluctuations that are time reversible, which suggests that fluctuations monitored during ageing are reversible \textit{if regarded as a function of the material time} instead of the laboratory time. Here ``reversible'' means that past and future cannot be distinguished in the statistical properties. This conjecture is investigated in this article. To do so it is necessary to both measure thermal fluctuations and determine the material time during the ageing. 

Our main experimental technique is dynamic light scattering (DLS) \cite{berne}. Compared to other experimental methods used for probing ageing dynamics (for example, dielectric spectroscopy) the DLS set-up operates in the time domain. With few exceptions \cite{Riechers2019} frequency-domain experiments average over several sinusoidal cycles, during which the ageing material may change properties. In contrast, time-domain experiments can, at least in principle, access ''instantaneous'' autocorrelation or response functions that are well defined even if ageing and correlation decay take place on comparable time scales. To determine the material time as a function of the time $t$, $\xi(t)$, we assume in the TN spirit that the normalized scattered-light intensity time-autocorrelation function is determined by the material-time difference $\xi(t_2)-\xi(t_1)$. As shown below, this leads to predictions that were discussed for spin-glasses in a seminal paper by Cugliandolo and Kurchan from 1994 \cite{cug94} - a paper, however, that did not discuss how an ageing system approaches thermal equilibrium.

It is an experimental challenge to obtain reliable thermal-fluctuation time-autocorrelation data during ageing because one cannot use the standard procedure of ensemble averageing by a moving time average. We determine the ensemble average for an ageing sample by simultaneously measuring several time-autocorrelation functions with a camera monitoring thousands of so-called speckles, that is, the granular interference pattern of the scattered light (Fig.\,\ref{fig1}b). From these data we extract the material time. Using standard methods of time-series analysis it is then demonstrated that, when regarded as a function of the material time, the fluctuations are stationary and reversible. These findings for the molecular system 1-phenyl-1-propanol (1P1P) apply also for the colloidal synthetic clay Laponite studied by VV (polarized) light scattering which, in contrast to the molecular system, does not approach equilibrium within the time of observation. In search for a counterexample we also studied a curing epoxy that ages chemically by continuous polymerization. However, this system also exhibits material-time reversal invariance; apparently, this property is not limited to physically ageing systems. Finally, material-time reversibility is shown to apply also for the potential-energy fluctuations of an ageing computer-simulated liquid.

We use depolarized DLS to probe the temporal intensity fluctuations of light scattered at 90$^\circ$ from the ageing sample. These fluctuations reflect the rotational dynamics of the molecules \textit{on the molecular scale} (see the SI for details), which is quantified by the intensity time-autocorrelation function. In standard equilibrium DLS experiments the latter is determined via a moving time average of the product of two intensities being separated by a fixed time interval $\Delta t$, but as mentioned this cannot be done for an ageing sample. To obtain acceptable statistics we perform a \textit{multispeckle} DLS experiment that probes a large speckle pattern using an sCMOS-camera with ten pictures taken per second; compare Fig.\,\ref{fig1}a. The intensity fluctuations of different speckles (Fig.\,\ref{fig1}c) are statistically independent, so the time-resolved ensemble-averaged intensity time-autocorrelation function $\langle I(t)\,I(t+\Delta t)\rangle$ can be determined by averageing over all pixels of the speckle pattern. This is the main idea; more details are given in the Methods section.

Time-resolved multispeckle light-scattering has been applied to monitor ageing and ultraslow dynamics of colloidal glasses, gels and foams, mainly in the diffusing-wave regime \cite{via02,cip03,kal05,Qi2017,aim18}, but so far this technique has not been used to monitor non-equilibrium molecular systems. The procedure of calculating multipixel averages to obtain time-resolved autocorrelations has also been used in connection with X-ray photon correlation spectroscopy, for example, for studying the physical ageing of metallic glasses \cite{rut12,eve15,cor23}.

If the intensity of the scattered light at time $t$ at one pixel is denoted by $I(t)$, the normalized intensity time-autocorrelation function is defined by $C(t_1,t_2)\equiv \langle I(t_1)I(t_2)\rangle/(\langle I(t_1) \rangle \langle I(t_2) \rangle)-1$ in which the angular brackets are averages over the different speckles. The normalized intensity time-autocorrelation function $C(t,t+\Delta t)$ of supercooled 1P1P \textit{in equilibrium} is plotted in Fig.~1d as functions of $\Delta t$ for different times $t$ (different colors). The data collapse onto a single curve, confirming the well-known stationarity of the equilibrium state. In contrast, a distinct $t$ dependence of $C(t,t+\Delta t)$ is observed following a temperature down jump as shown in Fig.~1e, where darker colors correspond to larger $t$. Starting from the initial equilibrium state at 195 K (left dashed curve) the rate of decay of $C(t,t+\Delta t)$ to zero as $\Delta t\to\infty$ slows down with increasing $t$. Eventually a new stationary equilibrium state is approached at the annealing temperature 193\,K (right dashed curve).

\section{Identifying the material time}

We determine the material time $\xi$ as a function of time from the assumption that the normalized intensity time-autocorrelation function $C(t_1,t_2)$ reflects the elapsed material time between $t_1$ and $t_2$. Thus, it is assumed \cite{avi13} that for some monotonous function that goes to zero for $x\to\infty$, $F(x)$, one has

\be\label{C_xi_eq}
    C(t_1,t_2)=F\bigl(\xi(t_2)-\xi(t_1)\bigr)\,.
\ee
For this to apply for any $t_1$ and $t_2$, the following consistency requirement must be obeyed. Since $\xi(t_3)-\xi(t_1)=\xi(t_2)-\xi(t_1)+\xi(t_3)-\xi(t_2)$, $C(t_1,t_3)$ is determined by $C(t_1,t_2)$ and $C(t_2,t_3)$ \cite{dou22}. This property always applies in thermal equilibrium because in that case $C(t_1,t_2)$ is a function of $t_2-t_1$, which is proportional to $\xi(t_2)-\xi(t_1)$ since $\xi$ in equilibrium is a linear function of the time $t$. During ageing, however, the consistency requirement is nontrivial as $\xi$ is no longer a linear function of $t$ because the structure changes with time. 

Introducing the abbreviated notation $C_{12}=C(t_1,t_2)$ etc, the above translates into the so-called triangular relation that was originally derived by Cugliandolo and Kurchan in 1994 in a mean-field calculation of time-autocorrelation functions of infinite-range spin systems quenched from infinite to a low temperature \cite{cug94},

\be\label{triang}
C_{13}\,=\,C_{13}(C_{12},C_{23})\,.
\ee
This equation expresses that, for any $t_2$, $C_{12}$ and $C_{23}$ determine $C_{13}$.  Equilibrium dynamics is time reversible, which implies symmetry of the function $C_{13}$,

\be\label{sym}
C_{13}(C_{12},C_{23})\,=\,C_{13}(C_{23},C_{12})\,.
\ee
This symmetry applies also for an ageing sample, a result that was derived by a different argument in Ref.~\onlinecite{cug94}. 

\begin{figure}
    \centering
    \includegraphics[width=70mm]{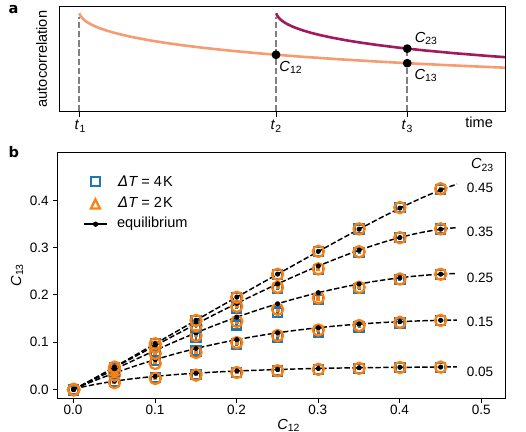}
    \caption{\textbf{Verifying the triangular relation.} (a) Schematic illustration how $C_{12}\equiv C(t_1,t_2)$, $C_{13}$, and $C_{23}$ are defined for $t_1<t_2<t_3$. (b) Average $C_{13}$ plotted vs. $C_{12}$ for fixed values of $C_{23}$ (1P1P data).
    The data refer to thermal equilibrium at 193\,K, as well as to physical ageing following two different temperature down jumps to 193\,K. The relation between the three time-autocorrelation functions for the two temperature jumps is the same as in equilibrium. The standard deviation of the $C_{13}$ data ($n=2.8561\cdot10^7$) are given as error bars (smaller than the symbol size, $<0.01$). These data confirm the triangular relation \eq{triang}, which expresses a necessary requirement for using the intensity time-autocorrelation function to define the material time via \eq{xi_C_eq}.}
    \label{fig2}
\end{figure}

We checked \eq{triang} by calculating $C_{12}$, $C_{23}$, and $C_{13}$ for several million time triplets, $t_1<t_2<t_3$, as illustrated in \fig{fig2}a. By binning subsets with same $C_{12}$ and $C_{23}$, we calculated the average $C_{13}$ and its standard deviation. In \fig{fig2}b, the average value of $C_{13}$ is plotted as a function of $C_{12}$ for fixed values of $C_{23}$ for 1P1P data measured in equilibrium, as well as after temperature jumps of different magnitude. All data are given with standard deviations as error bars that are so small that they are hardly visible (see SI). The small standard deviations and the collapse of data from different temperature protocols demonstrate that $C_{13}$ is determined by $C_{12}$ and $C_{23}$, independent of the temperature protocol. Thus, \fig{fig2}b validates the triangular relation in and out of equilibrium.

\begin{figure}
    \centering    
    \includegraphics[width=180mm]{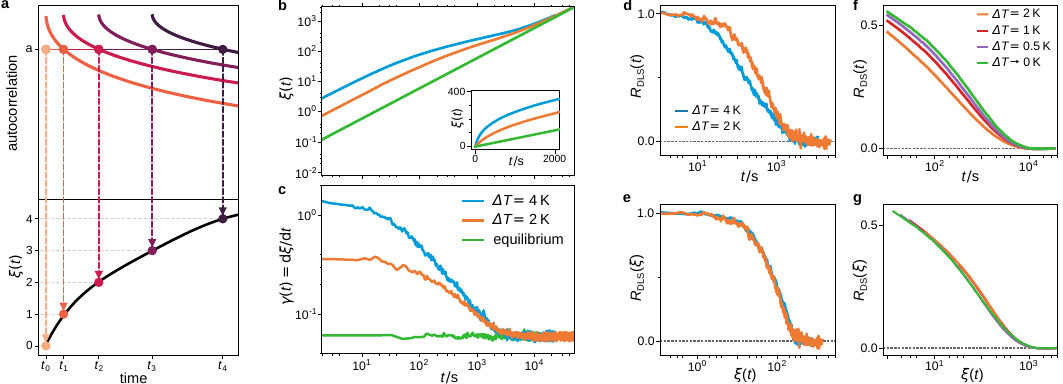}
    \caption{\textbf{Determining and validating the material time for down jumps to 193\,K.} 
    (a) Schematic illustration of how $\xi$ is identified step-by-step as a function of the laboratory time $t$. The time-autocorrelations are here plotted on a linear time scale, in contrast to the logarithmic time scale used in \fig{fig1}.
    (b) $\xi(t)$ in a logarithmic representation for different temperature jumps and in thermal equilibrium. At short times the samples have only aged insignificantly and $\xi(t)$ is a linear function of time; the same linear dependence applies in equilibrium that is approached for $t\to\infty$. The inset shows the same data in a linear representation (the 4\,K data were adapted to compensate for a slightly different annealing temperature, see the Supplement).
    (c) The ageing rate $\gamma$ (\eq{gamma}) as a function of $t$. Initially, $\gamma$ is large because the ``material-time clock'' ticks fast; as the ageing proceeds $\gamma$ approaches its equilibrium value. 
    (d) When considered as a function of the laboratory time, due to the fact that ageing is nonlinear the normalized relaxation functions (\eq{R}) differ in shape for the two temperature jumps. 
    (e) The transformation $t\to \xi(t)$ collapses the normalized relaxation functions.
    (f) Nonlinearity of physical ageing is evident also for the normalized relaxation functions of the dielectric loss measured at 10 kHz, $R_\mathrm{DS}(t)$. The green curve gives data for a temperature jump of 0.1 K, which to a good approximation is within the linear regime.
    (g) Plotting the dielectric $R_\mathrm{DS}(t)$ data as functions of the material time leads within the experimental uncertainty to the collapse predicted by the TN formalism. Note that the shapes of $R_\mathrm{DLS}(\xi)$ and $R_\mathrm{DS}(\xi)$ differ significantly.}
    \label{fig3}
\end{figure}

Having confirmed a necessary condition for defining the material time via \eq{C_xi_eq}, we note that inverting \eq{C_xi_eq} leads to for some function $\phi(x)$ 
\be\label{xi_C_eq}
    \xi(t_2)-\xi(t_1) = \phi\big(C_{12}\big)\,.
\ee
Thus, when the normalized intensity time-autocorrelation function has decayed to a certain value, a fixed amount of material time $\Delta\xi$ has elapsed. We use this to determine $\xi(t)$ in a step-by-step fashion, choosing the criterion $C(t_1,t_2)=a$ for $a=0.3$ to define $\xi(t_2)-\xi(t_1)=\Delta\xi\equiv 1$ (other choices of $a$ yield equivalent results, see the Supplement). The iterative procedure for determining the material time is illustrated in \fig{fig3}a using a linear time axis. \Fig{fig3}b shows $\xi(t)$ in a log-log representation for two temperature jumps (blue and orange), as well as  in equilibrium (green) where $\xi$ as mentioned is a linear function of $t$. In the inset $\xi(t)$ is plotted in a linear representation. \Fig{fig3}c shows the ageing rate defined by
\be\label{gamma}
\gamma\,\equiv\,\frac{\mathrm{d}\xi}{\mathrm{d}t}\,.
\ee
Following a temperature jump $\gamma$ goes from one value at the shortest times probed to the equilibrium rate at the annealing temperature. 

Next we validate the TN prediction that normalized relaxation functions of a given physical quantity following different temperature jumps are identical when regarded as functions of the material time $\xi$ \cite{nar71,scherer}. In order to do this, we again consider the time-resolved correlation function of the intensity fluctuations, $C(t,t+\Delta t)$, this time however as a function of the time $t$ evaluated at a constant, short lagtime (here $\Delta t=6\,$s). Writing in brief $C_{\Delta t}(t)$ and noting that if $\Delta t$ is chosen in a suitable manner, $C_{\Delta t}(t)$ represents the change of the fluctuations during ageing. In order to obtain a normalized relaxation function $R_\mathrm{DLS}(t)$ from $C_{\Delta t}(t)$ we define:
\be\label{R}
R_\mathrm{DLS}(t)\equiv\frac{ C_{\Delta t}(t)-C_{\Delta t}(t\rightarrow\infty)}{C_{\Delta t}(t=0)-C_{\Delta t}(t\rightarrow\infty)}\,.
\ee
\Fig{fig3}d shows $R_\mathrm{DLS}(t)$ as a function of time following 2\,K and 4\,K temperature down jumps, while \fig{fig3}e plots the same data as functions of the material time. The data collapse nicely in (e), confirming the TN prediction.

Going one step further, if the TN concept is genuinely valid, the material time determined using the time evolution of any material property during ageing should collapse the relaxation function constructed from any observable monitored during the ageing. To check this prediction, \fig{fig3}f shows temperature-jump data for a normalized relaxation function $R_\mathrm{DS}(t)$ constructed from dielectric loss data $\varepsilon''(\nu,t)$ obtained at the frequency $\nu=10\,$kHz, while (g) shows the same data as a function of the material time $\xi$ for the same temperature protocols as in the light-scattering experiment. As seen in the figure, there is an almost perfect collapse of the data. Considering that the dielectric data were taken in a different laboratory, the observed small deviations are well within the experimental uncertainties. The overall conclusion of panels (d) to (g) is that the TN material time identified from the light-scattering intensity time-autocorrelation function leads to a linear behavior when the relaxation during ageing is considered as a function of the material time, independent of the particular observable used to monitor the ageing process. This confirms that the material time has been identified correctly.

\section{Material-time stationarity and reversibility}

\begin{figure}
    \centering
    \includegraphics[width=180mm]{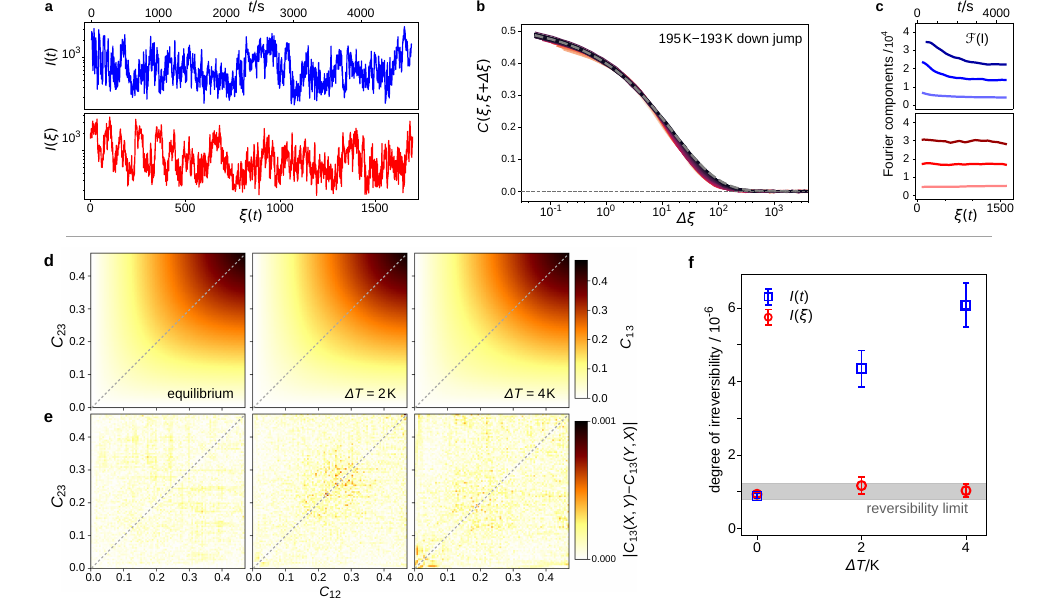}
    \caption{\textbf{Stationarity and reversibility of the intensity fluctuations regarded as functions of the material time.} 
    (a) Light-scattering intensity in one exemplary speckle for a $\Delta T=4\,$K down jump analyzed at equally spaced time intervals (blue), respectively, material-time intervals (red). 
    (b) shows the time-autocorrelation functions of Fig.~1e parameterized by the starting material time $\xi(t)$, plotted as a function of $\Delta\xi(t)$. The fact that the data collapse onto one curve shows that fluctuations during physical ageing are stationary functions of $\xi$.
    (c) Moving time-window Fourier components of the intensity at three frequencies plotted as a function of time (upper panel) and material time (lower panel). A systematic time dependence is observed, while the Fourier components are virtually constant as functions of the material time, confirming stationarity. 
    (d) shows the normalized intensity time-autocorrelation function $C_{13}$ indicated by the colors for given values of $C_{12}$ and $C_{23}$. Data are given for equilibrium and two temperature down jumps to 193\,K. The diagonal symmetry confirms \eq{sym}.
    (e) quantifies the deviation from symmetry, $|C_{13}(X,Y)-C_{13}(Y,X)|$, which is in all cases very small.
    (f) compares the degree of irreversibility for the intensity time series considered as functions of $t$ and $\xi$, quantified by the Jensen-Shannon divergence of the average visibility-graph Degree distributions \cite{Lacasa2008,Lacasa2012} averaged over $10^4$ time series. Each point reflects the mean value $\pm$SD of $n=10$ statistically independent analyses. The larger this quantity, the more irreversible is the time series in question. Note that zero can only reached for an infinitely long reversible time series; the gray  ``reversibility limit'' was identified by surrogate data testing \cite{Schreiber2000} performing the same analysis on a set of artificially generated reversible time series of the same statistics and length as the actual data time series (Methods).}
    \label{fig4}
\end{figure}

Having shown how to determine the material time during physical ageing from the normalized DLS intensity time-autocorrelation function, we now turn to this paper's central question: whether the linearization obtained by replacing time with material time translate into statistical reversibility of the intensity fluctuations.

Stationarity of a stochastic process is a necessary condition for reversibility \cite{law91}. The three top panels of \fig{fig4} investigate to which degree the intensity fluctuations are stationary when regarded as functions of the material time for a jump from equilibrium at 197\,K to equilibrium at 193\,K. \Fig{fig4}a shows the intensity data probed in one exemplary speckle as a function of $t$ (blue) and $\xi$ (red). Although the figures differ visibly, it is not immediately clear whether the lower data are more stationary than the upper. To investigate this we consider in \fig{fig4}b the material-time based correlation function of the speckle-averaged fluctuations, $C(\xi,\xi+\Delta\xi)$. This is found to be independent of $\xi$, meaning that $C(\xi,\xi+\Delta\xi)$ is material-time-translation invariant and thus stationary. \Fig{fig4}c illustrates the same aspect in the frequency domain by determining the Fourier components at three frequencies (for fixed-length moving time resp.\ material-time windows) and demonstrates that these become constant as functions of the material time (full spectrograms are shown in the SI).

After validating stationarity of the $I(\xi)$ time series, the lower panels of \fig{fig4} investigate whether reversibility applies in the sense that the time series of intensity fluctuations has the same statistics as that of the time-reversed series. \Fig{fig4}d shows data for the normalized intensity time-autocorrelation function in equilibrium and for physical ageing in a heat map showing $C_{13}\equiv C(t_1,t_3)$ (defined by the colors) as a function of $C_{12}$ and  $C_{23}$. The diagonal symmetry confirms the reversibility condition \eq{sym}. The degree of deviations from symmetry for given values of $X=C_{12}$ and $Y=C_{23}$, $|C_{13}(X,Y)-C_{13}(Y,X)|$, is investigated in (e) that shows deviations smaller than 1 promille.
A different test of material-time reversibility is given in (f), where we apply the visibility-graph algorithm \cite{Lacasa2008} to the intensity time series. This algorithm is well suited for our data as it works scale-free and assesses ''global'' aspects of reversibility \cite{Lacasa2012}. The idea is to map the time series onto a directed graph with nodes representing the time-series element edges defined according to a geometric criterion, \textit{in casu} visibility in the sense of how far one can ``see'' along the time axis \cite{Lacasa2008}. If a given time series is reversible, the visibility graph has the same statistical properties as the time-reversed one. This can be tested by comparing their respective Degree distributions using an appropriate statistical measure, e.g., the Jensen-Shannon divergence \cite{men97}. \Fig{fig4}f shows this measure of the degree of irreversibility for the $I(t)$ and $I(\xi)$ fluctuations (blue and red, respectively) where $\Delta T=0$ gives the equilibrium data. The gray area marks the lower range of values that can be obtained for a finite time series, as only for an \textit{infinite} reversible time series one can obtain a zero Jensen-Shannon divergence. We conclude that, within the experimental resolution, the physical ageing of 1P1P leads to scattered-light intensity fluctuations that are statistically reversible when regarded as functions of the material time, while they are clearly irreversible as functions of time.

\begin{figure}
    \centering
    \includegraphics[width=180mm]{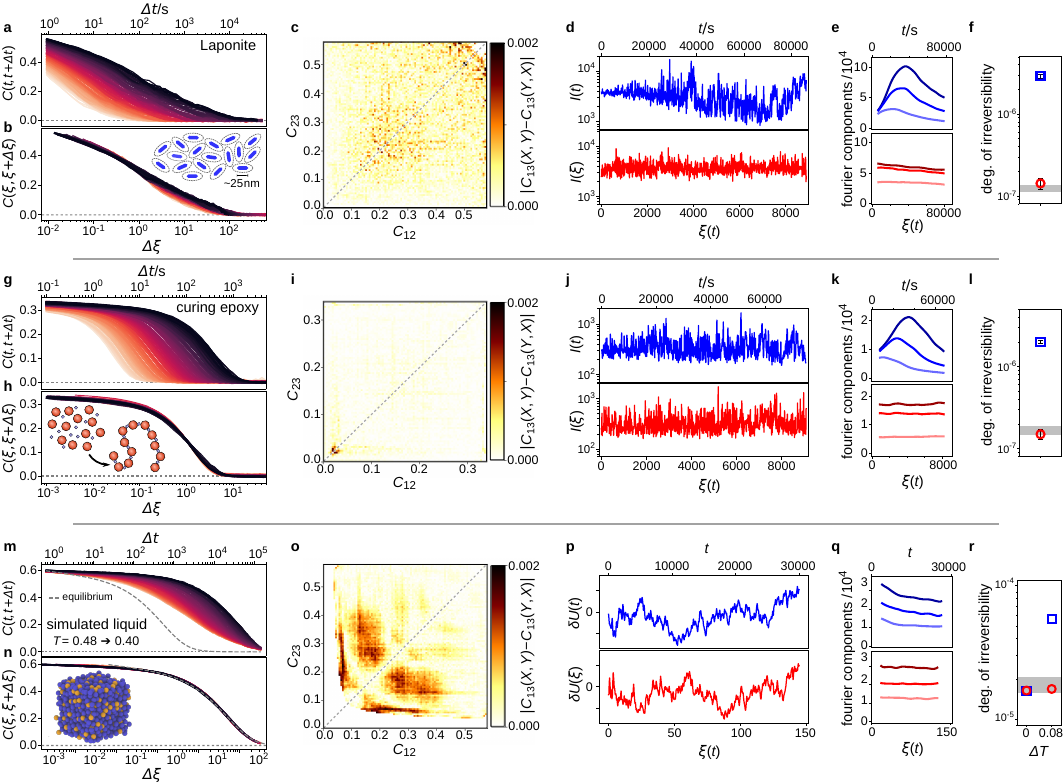}
    \caption{\textbf{Physical ageing data for the colloidal glass-former Laponite, chemical ageing data for a linear polymerizing epoxy, and computer simulation data for a binary Lennard-Jones system.}
    The top panels show data for disc-shaped Laponite particles dissolved in deionized water that gradually solidify into a colloidal glass. The center panels present the same analysis for the linear polymerization of bisphenol A diglycidyl ether initiated upon mixing with appropriate linker molecules. The bottom panels show ageing of the potential-energy fluctuations of a binary Lennard-Jones mixture after a temperature jump from 0.48 to 0.40 (dimensionless units). (a)/(g)/(m) show the slow down of time-autocorrelation functions as the annealing time increases. (c)/(i)/(o) show heat maps analogous to those of \fig{fig4}e confirming material-time reversal invariance. The necessary condition of stationarity of the intensity time series, shown in (d)/(j)/(p), with regard to the material time (red) is confirmed for the corresponding Fourier components in (e)/(k)/(q) and for the time-autocorrelation functions in (b)/(h)/(n). (f)/(l)/(r) show results analogous to that of \fig{fig4}f. In panels (f) and (l), each point reflects the mean value  $\pm$SD of $n=10$ statistically independent analyses. Note that (r) also includes data for an equilibrium simulation ($\Delta T=0$). }
    \label{fig5}
\end{figure}

\section{Results for other systems} 

To test the generality of the above findings we also investigated Laponite, a colloidal system of synthetic 25\,nm sized disc-shaped clay particles that after stirring in water solidifies gradually into a transparent colloidal glass \cite{ruz11}. In this case one cannot probe the full approach to equilibrium, because the system keeps solidifying within any realistic time window of observation. Nevertheless, following the above procedure one can define a material time from the normalized scattered-light intensity time-autocorrelation function. In order to vary the experimental conditions, we changed from depolarized (VH) to polarized (VV) light-scattering geometry, which probes local particle-concentration fluctuations \cite{berne}.

The results for Laponite are summarized in the top panels of \fig{fig5} in which (a) shows data for the normalized intensity time-autocorrelation function. Clearly, the particle dynamics slows down considerably during the experiment. Material-time stationarity is validated in \fig{fig5}b, while (c) confirms the symmetry of $C_{13}(C_{23},C_{12})$ analogous to \fig{fig4}e, thus demonstrating material-time reversibility. There is not a perfect data collapse in (b), but it should be noted that these data cover more than two decades of relaxation times. Possibly, a fast microscopic process \cite{Marques2015} or a decoupling of translational and rotational degrees of freedom \cite{Jabbari2012} could lead to slight deviations from perfect stationarity. \Fig{fig5}c shows intensity time and material-time series from one exemplary speckle. The amplitudes of the fluctuations as functions of $t$ and $\xi$ differ visibly as a consequence of the averageing over exposure time intervals that are fixed in regard to $t$ or $\xi$, respectively, see SI for a detailed discussion of this. The effect is more pronounced for the Laponite data due to the large change of the relaxation time during ageing. Finally, (e) shows three of the respective Fourier components evaluated over moving time windows, while (f) demonstrates material-time reversibility following the same procedure as in \fig{fig4}f. 

Our work set out to investigate whether material-time reversibility is a characteristic of \textit{physical} ageing. Having confirmed this for two quite different systems, we searched for a counterexample and decided to investigate an irreversibly reacting chemical system, namely the linearly polymerizing epoxy based on bisphenol A diglycidyl ether (see Methods section). We monitored the dynamics during polymerization by VH depolarized light scattering. The results are shown in the middle panels of \fig{fig5}, which are analogous to those of Laponite. Surprisingly, no counterexample to material-time reversibility is provided by this chemically ageing system; in fact the findings are very similar to those of physically ageing systems by demonstrating stationarity in (k), collapse of correlation functions in (h), and material-time reversibility in (i) and (l). 

Finally, we simulated a modification of the Kob-Andersen binary Lennard-Jones (LJ) model of a supercooled liquid \cite{Kob1995,Toxvaerd2009} following a temperature down jump from a state of equilibrium. The quantity monitored during ageing is the instantaneous value of the potential energy of each particle (see Methods section). The same analysis as for the experimental data was carried out for the per-particle potential energy, demonstrating the generality of our findings in \Fig{fig5}(m--r). The single-particle potential-energy fluctuation time-autocorrelation functions are shown in (m) as a function of time. They collapse almost perfectly as functions of the material time in (n), with small deviations appearing only at short $\Delta\xi$. These deviations are due to additional fast processes that appear at even shorter time scales and are virtually unaffected by ageing. Such fast processes are found in almost all glassy materials, but are usually much further separated from the structural relaxation and thus not noticed in most ageing experiments. Material-time reversibility is tested in (o), where again minor deviations from perfect symmetry are observed. For further testing reversibility we studied time-series reflecting the summed potential energy fluctuations of 200 random particles, $\delta U$. The resulting degrees of irreversibility are shown in (r), where $\delta U(\xi)$ is shown to be just as reversible as $\delta U(t)$ of the equilibrium liquid at the annealing temperature. This result was obtained after applying a narrow Gaussian filter to eliminate the fast non-material-time stationary/reversible microscopic fluctuations, which otherwise lead to subtle residual irreversibility in $\delta U(\xi)$, see Methods and SI.

\section{Discussion}
We have presented ageing data for a supercooled molecular liquid obtained by multi-speckle dynamic light scattering. By averageing over a multitude of speckles, time-resolved intensity time-autocorrelation functions can be determined with good accuracy during ageing. This presents a significant advantage over other experimental approaches used to monitor the dynamics of ageing. Our approach allows a detailed analysis of the shape of time-autocorrelation functions during ageing, also when the time scales of ageing and the observed correlation decay are similar.  

We used the light-scattering data to validate the material-time concept of the TN formalism. This concept, which has been applied successfully for half a century to describe physical ageing \cite{nar71,scherer,mck94,mck17}, is remarkable by transforming a highly nonlinear phenomenon into a linear one simply by replacing time with material time. Since standard linear-response theory reflects the time reversibility of the fundamental laws of physics, this suggests that fluctuations during physical ageing are statistically reversible when regarded as functions of the material time. We validated this for light-scattering data on three ageing systems: 1-phenyl-1-propanol (1P1P), Laponite, and a curing epoxy -- as well as for data on a computer-simulated ageing LJ model liquid. Note that these systems differ in several respects: 1P1P and epoxy are molecular while Laponite is colloidal; 1P1P, Laponite and LJ age physically while epoxy ages chemically; Laponite and epoxy do not converge to equilibrium within the window of observation while 1P1P and LJ do. Moreover, our experiments probed dynamics on different length scales: while reorientation of the optical anisotropy on a molecular scale is probed in VH light scattering of 1P1P, in the case of VV scattering on Laponite density fluctuations are probed on the scale of the optical wavelength. In VH scattering of the curing epoxy, it is mainly hardener/resin concentration fluctuation effects that are probed. Finally, similar to 1P1P, the computer simulations access the single-particle scale.
Despite these significant differences, all four systems obey the triangular relation \eq{triang} and have fluctuations that within the experimental resolution are stationary and reversible when considered as functions of the material time.

It is important to emphasize that while we have demonstrated material-time reversibility of the light-scattering intensity statistics on the time scale of the structural relaxation, as is evident from the simulations on the LJ system the description in terms of a material time does not lead to reversibility of the microscopic dynamics in the picosecond range. Moreover, even when the light-scattering intensity fluctuations are material-time reversible, other quantities may well show irreversible behavior. This \textit{caveat} has recently been discussed by O'Byrne and coworkers in connection with the dynamics of active matter \cite{byr22} demonstrating that, while the microscopic dynamics of active matter is not time-reversible, the dynamics of observables referring to the coarse-grained level may appear so. Since establishing absolute reversibility involves testing infinitely many different variables, it cannot be excluded that our ageing samples are irreversible at some deeper level. To investigate this in the future, a promising approach would be to analyze light-scattering and other data using machine-learning techniques, as recently suggested by Seif \textit{et al.} \cite{sei21}.

In view of our findings the following question arises: Does the material time determined by one specific method, say DLS, uniquely determine the transient state of the ageing system \cite{scherer,mck17}?  That is, do all other observables probing ageing become linear and the corresponding fluctuations become stationary and time reversible when considered as a function of the material time? We made a first attempt to elucidate these questions by showing that relaxation functions constructed from dielectric data indeed do become linear using the DLS material time (\fig{fig3}g). Whether this holds in general and includes other static and dynamic observables remains an open question \cite{Peredo2022}. We note that results from different linear-response experiments show that, at least for some glass-forming liquids, the ratio of different relaxation times is independent of the temperature \cite{roe21}, which is consistent with the existence of a unique material time.

While the concept of reversibility is fundamental in physics, in the present context we considered it more generally as a formal time-series property as, e.g., recently discussed in connection with neuroscience \cite{kri23}. We are not aware of a theoretical framework that predicts material-time statistical reversibility of fluctuations during ageing. 
The TN ageing description is phenomenological and has no predictions for fluctuations. Assuming in the TN spirit that the intensity time-autocorrelation function during ageing is a function of the material-time increase, the triangular relation \eq{triang} and the symmetry relation \eq{sym} were derived above; in that approach these identities may be regarded as inherited from thermal equilibrium. \eq{triang} and \eq{sym} were originally derived, however, by Cugliandolo and Kurchan (CK) in an entirely different context \cite{cug94,cha07}. CK studied spin models with infinite-range interactions and showed that the exact Schwinger-Dyson equations for fluctuation and response imply \eq{triang} and (\ref{sym}) at low temperatures where the system never converges to equilibrium. At higher temperatures, on the other hand, the system converges to equilibrium and no triangular relation was predicted to apply. Under the latter conditions, however, we do find that the triangular relation applies, as shown above for 1P1P and the computer simulations of simple Lennard-Jones type systems ageing to equilibrium \cite{avi13,dou22}. The CK mean-field approach is exact in infinite dimensions \cite{cug94,ago19}. Although the basic CK predictions were later proposed to apply also in finite dimensions \cite{cas07,avi13}, the range of applicability of the mean-field scenario remains unclear \cite{arc20}. Interestingly, the subject of time-reparametrization invariance pioneered by CK \cite{cug94}, a consequence of the triangular relation, has recently become fashionable in field theory where it represents the ``gravity'' low-energy limit \cite{fac19,kur23}.

Our results give rise to a number of questions: Is material-time reversibility a universal characteristic of ageing? Is it possible to formulate generalized fluctuation-dissipation relations based on material time instead of laboratory time, possibly on a coarse-grained level? If so, how would this relate to the effective-temperature concept introduced long ago for the description of ageing \cite{cug11}? Does material-time reversibility reflect the reversibility of the fundamental laws of physics and thereby connect to the fluctuation theorem that quantifies a consequence of microscopic time reversibility \cite{boc81,eva02}? -- To illuminate these questions it would be interesting to find an ageing system that violates material-time reversibility, but we have so far failed to do so. Possibly, this could be achieved by performing very large temperature jumps, as suggested by recent results \cite{Canglialosi2013, Herrero2023}.

\section*{Corresponding authors}
\noindent Till Böhmer (till.boehmer@pkm.tu-darmstadt.de)\\
Jeppe C. Dyre (dyre@ruc.dk)\\
Thomas Blochowicz (thomas.blochowicz@physik.tu-darmstadt.de)

\section*{Author contribution statement}
J.C.D and Th.B. devised the project; 
T.B and J.P.G planned the experiments; 
T.B, J.P.G and J.-N.K performed the experiments; 
T.B. and J.P.G. analyzed the experimental data; 
L.C. and T.B. performed and analyzed the computer simulations; 
T.H., J.C.D. and Th.B. supervised the experiments and data analysis, 
T.B. and J.C.D. wrote the manuscript with input from all authors;
J.C.D. and Th.B. provided resources and acquired funding.

\section*{Competing Interests Statement}
The authors declare no competing interests.

\acknowledgments{The authors thank Massimiliano Zanin and Susanne Ditlevsen for advice regarding the data analysis. This work was supported by the VILLUM Foundation's \textit{Matter} grant (VIL16515) and by the Deutsche Forschungsgemeinschaft (grants no. BL 923/1 and BL 1192/3).}

\section*{Methods: Experimental procedure and data analysis}

\subsection{Multispeckle DLS setup and temperature protocols}
Experiments are performed in a custom-built light-scattering setup, where a Cobolt Samba 500 Nd:YAG laser ($\lambda=532$\,nm) is used to illuminate the sample with vertically polarized light. The sample cell is mounted inside a Cryovac cold-finger cryostat surrounded by a high vacuum generated by a vibration-free Agilent ion getter pump. Scattered light is detected at $90^\circ$ angle using a Hamamatsu ORCA-Flash 4.0 V2 sCMOS camera. Light is guided onto the camera chip by a custom-built optical system consisting of a circular aperture, a $f=6\,$cm spherical lens, a fine-adjustable Glan-Thompson polarizer from B. Halle with extinction ratio $10^{-6}$, an adjustable slit aperture, and a 2\,nm band-pass filter (in the respective order toward the detection unit).

To calibrate the temperature protocols, the sample temperature must be monitored precisely. This has been achieved by constructing a dummy sample cell of the same geometry as the original light-scattering cell, but equipped with two PT100 temperature sensors: The first one is located inside the liquid sample, the second one is glued into a channel drilled through the aluminum bottom of the sample cell using Loctite Stycast 2850FT thermally conducting epoxy glue. During the light-scattering measurements only the second sensor could be used, as positioning a sensor inside the liquid interferes with the laser beam. Using the dummy sample cell it was confirmed that both sensors coincide for different temperature protocols, implying that the temperature determined by the second sensor is an appropriate measure of the actual sample temperature.

To perform a fast temperature down jump from the starting temperatures $T_0=197\,\mathrm{K},\,195\,\mathrm{K}$ to the ``annealing'' temperature $T_\infty=193\,$K with amplitude $\Delta T = T_0-T_\infty$ it is not sufficient to simply quench the cold-finger from $T_0$ to $T_\infty$ and let the sample equilibrate there. Due to the thermal lag between cold-finger and sample cell, the sample temperature would initially decrease quickly, but as soon as the cold-finger reaches $T_\infty$ the cooling rate inside the sample declines and it takes more than $1000\,$s for the sample temperature to reach $T_\infty$. To mitigate this we initially cooled the cold-finger several degrees below $T_\infty$ as quickly as possible. This temperature was then held for a short time $t$, after which the cold finger was heated back to $T_\infty$ with heating rate $c$. By choosing $t$ and $c$ in an optimal way, it was possible to have the sample temperature quickly reach $T_\infty$ without any significant temperature undershoot (see SI for details). The parameters for the  optimal temperature protocol were determined for $\Delta T=$0.5, 1.0, 2.0 and 4.0\,K prior to the light-scattering measurements (0.5 and 1.0\,K are not shown, but used for the analysis in Fig. 3g). We limit the study to these temperature jump amplitudes because larger temperature jumps would require larger cooling rates that cannot be obtained with the present setup. During the measurements, the prepared temperature protocols were repeated, and reproducibility was ensured by controlling the Lakeshore 335 temperature controller via Python scripts. Following this procedure, temperature jumps could be performed in less than $200\,$s with an accuracy of 0.1\,K. The final temperature could then be held for $>$24\,h with an accuracy of $\pm 0.05$\,K.

\subsection{Sample preparation and measurements}
\subsubsection{1P1P}
1-phenyl-1-propanol (1P1P; Alfa Aesar, 98+\% purity) was filtered into the light-scattering sample cell using a 450\,nm syringe filter. The sample was first kept at the starting temperature $T_0$ for 12\,h to ensure the supercooled liquid is in (metastable) thermal equilibrium. A camera measurement quantifying the equilibrium state at $T_0$ was performed with exposure time $0.02\,$s over 300,000 frames (6000\,s). After this, a second camera measurement was started with an exposure time of $0.2\,$s over 300,000 frames ($\sim 17$\,h). Simultaneously, the prepared temperature protocol was initiated. The duration of the measurement was chosen to be significantly longer than the time needed for the sample to equilibrate, which allowed us to analyze in detail the equilibrium intensity fluctuations. All experiments involving 1P1P were performed in the VH (depolarized) geometry, thus the reorientation of the molecular optical anisotropy tensor is probed \cite{berne}.

For the dielectric measurements the complex capacity was measured at the fixed frequency $\nu_0=$10\,kHz, with an Andeen-Hagerling AH 2700A high-precision bridge in a custom-built cryostat system \cite{iga08a,iga08b}. Fast temperature jumps up to $\Delta T$ = 4\,K with microkelvin precision were obtained by a subcryostat system based on a Peltier element and a nonlinear temperature sensor \cite{iga08a,hec10}. The sample temperature can be changed in a few seconds. 

From the complex capacity, the complex permittivity of 1-phenyl-1-propanol at 10\,kHz was determined and its imaginary part, $\varepsilon^{\prime\prime}(t)$, was used to calculate the dielectric ageing function, $R_{\text{DS}}(t)$ (analogous to \eq{R})
\begin{equation}
    R_{\text{DS}}(t)=\frac{\varepsilon''(t)-\varepsilon''(t\rightarrow \infty)}{\varepsilon''(t=0)-\varepsilon''(t\rightarrow \infty)}\,.
\end{equation}
Here $t=0$ refers to the equilibrium value before the jump. The dielectric ageing function is thus normalized to the difference in the equilibrium values at the start and end temperatures for the temperature jump. Data were collected linearly in time and averaged using a logarithmic binning scheme. For combining DLS and DS measurements we account for slight differences regarding the annealing temperature, which inevitably appear due to the experiments being performed in different laboratories. We determine the respective relaxation time at the annealing temperatures of both experiments by considering the results from a detailed analysis of 1P1P in both methods in Ref.\,\cite{Boehmer2019} and apply the procedure presented in S1 of the Supplementary Information to consider the temperature difference.

\subsubsection{Laponite}
The Laponite suspension was prepared in consideration of the phase diagram\,\cite{Tanaka2004} with parameters chosen such that a colloidal Wigner glass was obtained: Laponite powder (Laponite-RD from BYK) was dried for one week at $1\,$mbar to remove all water. Water with pH 10 and ionic strength $I<10^{-4}\,$M was obtained by adding an appropriate amount of NaOH to milli-Q water. Subsequently, 2.98\,wt\% of Laponite was added and the mixture was stirred for 24\,h. Finally, the sample was filtered into a glass cylinder using a 450\,nm syringe filter, which may have led to a slight reduction of the Laponite concentration. Physical ageing starts directly after the mixture is filtered; however for the first few hours the colloidal dynamics is too fast to be captured by the camera. The camera measurement with exposure time 0.1\,s was started several minutes after filtration, and the scattered light was monitored in VV (polarized) geometry over 1,500,000 frames ($\sim 42\,$h). Laponite was studied at room temperature.

\subsubsection{Epoxy}
The epoxy system is based on bisphenol A diglycidyl ether resin (Alfa Aesar). As polymerization agent we used N,N,N',N'-tetraethyldiethylenetriamine (Sigma Aldrich, 90\%). This specific agent induces a linear polymerization; thus the mixture was prepared with a 1:1 molar ratio. Before mixing, the resin was dried and degassed in a vacuum oven. Afterwards, it was filled into a glass syringe equipped with a stainless steel filter holder containing a 450\,nm nylon membrane filter suitable for operation at high temperatures. The syringe was then heated to 150$\,^\circ$C to decrease the viscosity of the resin, which allowed us to filter the resin into a dust-free sample glass. The appropriate amount of hardener was added, again, using a syringe filter. Subsequently, the mixture was magnetically stirred at 400\,rpm for 10\,min and filled into a dust-free cylindrical glass sample holder. Air-bubbles were removed by exposing the mixture to vacuum for 15\,min. Finally, the glass tube was sealed and placed inside a suitable sample oven preheated to 310\,K, and the camera measurement in VH geometry with exposure time 0.1\,s was started for 2,000,000 frames ($\sim$56\,h)

\subsection{Molecular Dynamics}
The data for the binary Lennard-Jones (LJ) system presented in Fig. \ref{fig5} were produced using the GPU-optimized software RUMD \cite{RUMD}. The interaction potential used is the modification of the standard Kob-Andersen potential introduced in Ref. \cite{sch20} to counteract crystallization. Jumps from temperature $0.48$ to $0.40$ at density $1.20$ for a system of $N=8000$ particles ($6400$ A particles and $1600$ B particles) were simulated for $1.68\cdot10^5$ LJ time units with a time step of $0.005$. The simulations ran in the \textit{NVT} ensemble using a standard Nosé-Hoover thermostat with relaxation time of $0.2$ in LJ units. Velocities were re-scaled at the start of the simulation to temperature $0.40$ in order to avoid nonphysical behavior of the thermostat. The potential energy of each particle was saved every $128$ time steps (corresponding to 0.64 LJ time units). Simulations from $30$ independent starting configurations were run. The data shown in Fig. \ref{fig5} refer to the A particles only and consider a total of 192,000 particles in order to obtain time-resolved autocorrelations with minimal statistical noise. Starting configurations were obtained from an equilibrium simulation at density $1.20$ and temperature $0.48$, separated in time by $4.2\cdot10^4$ LJ units (roughly corresponding to $80$ relaxation times).

\subsection{Data analysis}
For all temperature jumps of 1P1P, for Laponite during annealing, and for the curing epoxy, the resulting camera movie files were analyzed using three different techniques to obtain time-resolved autocorrelation functions, pair-wise autocorrelations for time triplets, and intensity time series (TS) as a function of time $t$ and material time $\xi$. Similar procedures were used to analyze the simulation data. In the following, each procedure and subsequent analyses are explained.

\subsubsection{Time-resolved autocorrelation functions}
Time-resolved auto-correlation functions were obtained from the camera movie via a multi-pixel average over images consisting of $N=2048$ columns and $M=350$ rows. Due to the Gaussian intensity profile of the laser beam, the average pixel intensity varies slightly perpendicular to the propagation direction of the laser (resp. for different rows). At the same time, the average intensity is constant along the propagation direction (resp. for different columns). Thus, normalization of auto-correlations is carried out independently for each row. Finally, the result is averaged over all rows. Mathematically, our procedure amounts to

\be\label{trc}
    C(t,t+\Delta t) = \frac{N}{M}\sum\limits_{j=1}^{M} \left[  \frac{\sum_{i=1}^{N}\,I_{ij}(t)I_{ij}(t+\Delta t)}{\Bigl(\sum_{i=1}^{N}\,I_{ij}(t)\Bigr)\Bigl( \sum_{i=1}^{N}\,I_{ij}(t+\Delta t)\Bigr)}  \right] -1,
\ee
with column $i$ and row $j$. $C(t,t+\Delta t)$ was calculated for all available $t$ (each frame) and for 47 different logarithmically equally spaced values of $\Delta t$, where the smallest value is the exposure time of the camera.

From the simulation data, we calculate time-resolved autocorrelations of the single-particle potential energy fluctuations $\Delta u_i(t)=u_i(t)-\langle u_i(t)\rangle_N$, where $\langle...\rangle_N$ indicates average over all particles $i\in\{1,2,...,N\}$,

\be\label{simtrc}
    C(t,t+\Delta t) = \frac{\Bigl\langle \Delta u_i(t) \Delta u_i(t+\Delta t)\Bigr\rangle_N}{\Bigl\langle \bigl(\Delta u_i(t)\bigr)^2\Bigr\rangle_N \Bigl\langle \bigl(\Delta u_i(t+\Delta t)\bigr)^2\Bigr\rangle_N}.
\ee

\subsubsection{Pairwise autocorrelation for time triplets}
In order to verify the triangular relation \eq{triang}, we analyzed many time triplets $t_1<t_2<t_3$. The analysis focused on the time frame in which $\gamma(t)$ changes, as the triangular relation holds trivially when $\gamma(t)=\mathrm{const}$. In this time frame, $10^3$ linearly spaced values of $t_1$ were chosen. For each $t_1$, different $t_2=t_1+\tau^\prime$ were constructed for 169 logarithmically equally spaced values $\tau^\prime$. Subsequently, for each combination of $t_1$ and $t_2$, different $t_3 = t_2 + \tau^{\prime\prime}$ were constructed using the same 169 logarithmically equally spaced values for $\tau^{\prime\prime}$. Combined, this procedure generated $1000\cdot 169^2\cong 3\cdot 10^7$ different time triplets. For each of these, the pairwise auto-correlations $C_{12}$, $C_{13}$ and $C_{23}$ (see the main text for notation) were calculated using \eq{trc} with, e.g., $t=t_1$ and $t+\tau=t_2$ for the calculation of $C_{12}$. Finally, subsets with same values of $C_{12}$ and $C_{23}$ were binned (with resolution $\Delta C=0.005$), and for each subset the mean $\overline{C_{13}}(C_{12},C_{23})$ and corresponding standard deviation $\sigma_{13}(C_{12},C_{23})$ was calculated.
\\

\subsubsection{Constructions of the intensity time series (TS)}
To obtain the intensity TS in laboratory time $I(t)$, resp. material time $I(\xi)$, intensity fluctuations in one pixel were first interpolated at a tightly spaced sequence of times, where the elapsed time between consecutive points was chosen to be constant in time, 
resp. material time. As the focus is on physical ageing, the analysis is limited to the time frame where $\gamma(t)$ changes (same time frame as for time triplets). Afterwards, subsets were binned and averaged to obtain a TS of length $N\in\{500,1000,2500,5000,10000\}$. For $I(\xi)$, this procedure essentially simulates the result one would obtain using a camera operated with a time-dependent exposure time defined by the material-time clock rate. Different $N$ were analyzed to ensure the observed effects are universal in regard to temporal resolution. As qualitatively equivalent results were obtained for all $N$, only results for $N=2500$ are shown in the paper. The above described procedure was applied to $10^4$ different pixels chosen such that the fluctuations are statistically independent. The entire subset is then analyzed regarding stationarity and reversibility.

The procedure is very similar for the simulations, except that instead of binning we apply a Gaussian weighted time-average at times chosen to be constant in time, resp. material time. The width of the Gaussian filter is proportional to the time clock-rate (constant), resp.\ material-time clock-rate. This procedure gives very similar results to binning, but is able to filter out very fast fluctuations more efficiently (see the SI). This is necessary for the simulations because of the small dynamic separation between the fast microscopic process and structural relaxation (a consequence of the relatively high temperature at which simulations are performed). As discussed in the manuscript, the fast process does not age and does not conform to the material-time formalism.
\\

\subsubsection{TS stationarity analysis}
TS stationarity is tested by Fourier analyzing short sequences of the fluctuations to obtain time-resolved Fourier components. The corresponding procedure involves multiplying the original intensity TS by a Gaussian function for which we can vary the maximum position $t_0$ to obtain temporal resolution. The obtained "time-resolved" TS are analyzed using the Fast Fourier Transformation (FFT) algorithm, and the results are averaged over $10^4$ statistically independent TS. More details of this analysis can be found in the SI.

\subsubsection{TS time-reversibility analysis}
Time-reversibility of $I(t)$ and $I(\xi)$ was analyzed using the visibility graph (VG) algorithm \cite{Lacasa2008,Lacasa2012}. This was done by transforming a TS $I(t)$ resp. $I(\xi)$ and its respective time-inverted TS, $I^-(t)$ resp. $I^-(\xi)$, into a VG using the fast algorithm\,\cite{Xin2015} from the \textsc{ts2vg} python package. Subsequently, the Degree distributions were calculated. Finally, the Degree distributions of original and time-inverted TS, $P(k)$ and $P^-(k)$, were averaged over $10^4$ statistically independent TS. The difference of both distributions was quantified using the Jensen-Shannon divergence
\be
D_\mathrm{JS}\bigl(P(k)||P^-(k)\bigr) = \frac{1}{2}\Bigl(D_\mathrm{KL}\bigl(P(k)||\tilde{P}(k)\bigr)+D_\mathrm{KL}\bigl(P^-(k)||\tilde{P}(k)\bigr)\Bigr),
\ee
where $\tilde{P}(k)=\frac{1}{2}\bigl(P(k)+P^-(k)\bigr)$ and
\be
    D_\mathrm{KL}\bigl(P(k)||\tilde{P}(k)\bigr)=\sum\limits_k\,P(k)\log\frac{P(k)}{\tilde{P}(k)}
\ee
is the Kullback-Leibler divergence. $D_\mathrm{JS}$, which is a symmetric version of $D_\mathrm{KL}$, is referred to as the \textit{degree of irreversibility} in the paper.

Uncertainties for $D_\mathrm{JS}$ were determined by repeating the above described analysis ten times for different subsets of $10^4$ pixels (no pixel was used in more than one subset). The standard deviations of the thus obtained distributions of $D_\mathrm{JS}$ are plotted as error bars in Figs. 4 and 5. For the simulations, the uncertainties of $D_\mathrm{JS}$ could not be determined due to the limited number of analyzed particles.

Due to finite-size effects, $D_\mathrm{JS}>0$ even for a time-reversible TS. To obtain the lower limit of $D_\mathrm{JS}$ that would be approached if the experimentally observed TS were truly time-reversible, we performed a surrogate analysis \cite{Schreiber2000}. Numerous surrogate TS of $I(\xi)$ and $I^-(\xi)$ were calculated using the Iterated Amplitude Adjusted Fourier Transform (IAAFT) algorithm\cite{Schreiber1996} and analyzed in subsets of $10^4$ TS (equivalent to the procedure of analyzing experimental TS). The reversibility limits shown in Figs. 4 and 5 correspond to the distribution of $D_\mathrm{JS}$ obtained by performing several such surrogate analyses. 

\section*{Supplementary information contents}
Temperature protocol and handling slight temperature differences, confirmation that DDLS probes 1P1P dynamics on the molecular scale, influence of the parameter $a$ on the material time, standard deviation $\sigma_{13}$ in the triangular relation, effects of time-reparametrization on the amplitude of fluctuations, time-resolved Fourier analysis, filtering procedure for simulation TS data, and more details regarding the statistical analysis of reversibility.

\section*{Data availability}
The study generated $\sim10\,$TB of data, which are available from the corresponding authors upon request.

\section*{Code availability}
The code for the analysis of all data sets is available from the corresponding authors.

\end{document}